\documentclass{ws-ijmpd}
\usepackage[super,compress]{cite}
\usepackage{longtable}
\usepackage{amsfonts}
\usepackage{amsmath}
\usepackage{amssymb}
\usepackage{bm}
\usepackage{dcolumn}
\usepackage{epsfig}
\usepackage{graphicx}
\usepackage{graphics}
\usepackage[latin1]{inputenc}
\usepackage{latexsym}
\usepackage{rotating}
\usepackage{hyperref}
\usepackage{color}
\usepackage{soul}
\usepackage[normalem]{ulem}
\newcommand\be{\begin{equation}}
\newcommand\ba{\begin{eqnarray}}
\newcommand\ee{\end{equation}}
\newcommand\ea{\end{eqnarray}}

\begin{document}

%
\catchline{}{}{}{}{}
%

\title{Observational constraints on Gauss-Bonnet cosmology} 
\author{Micol Benetti} 

\address{Departamento de Astronomia, Observat\'orio Nacional, 20921-400, Rio de Janeiro, RJ, Brasil\\
{micolbenetti@on.br}}
\author{Simony Santos da Costa} 

\address{Departamento de Astronomia, Observat\'orio Nacional, 20921-400, Rio de Janeiro, RJ, Brasil \\
Dipartimento di Fisica "E. Pancini", Universit\'a di Napoli "Federico II", Via Cinthia, I-80126, Napoli, Italy\\
simonycosta@on.br}
\author{Salvatore Capozziello} 

\address{Dipartimento di Fisica "E. Pancini", Universit\'a di Napoli "Federico II", Via Cinthia, I-80126, Napoli, Italy \\
{Istituto Nazionale di Fisica Nucleare (INFN), Sez. di Napoli, Via Cinthia 9, I-80126 Napoli, Italy}
{Gran Sasso Science Institute, Via F. Crispi 7, I-67100, L' Aquila, Italy}
{Lepage Research Institute, Ul. 17. Novembra 1, 08116 Presov, Slovakia.}\\
capozzie@na.infn.it}

\author{Jailson S. Alcaniz} 

\address{{Departamento de Astronomia, Observat\'orio Nacional, 20921-400, Rio de Janeiro, RJ, Brasil}
\\Physics Department, McGill University, Montreal, QC, H3A 2T8, Canada\\
Universidade Federal do Rio Grande do Norte, 59072-970, Departamento de
F\'{\i}sica, Natal, RN, Brazil\\
alcaniz@on.br}

\author{Mariafelicia De Laurentis} 

\address{Dipartimento di Fisica "E. Pancini", Universit\'a di Napoli "Federico II", Via Cinthia, I-80126, Napoli, Italy \\
Istituto Nazionale di Fisica Nucleare (INFN), Sez. di Napoli, Via Cinthia 9, I-80126 Napoli, Italy\\
Institute for Theoretical Physics, Goethe University, Max-von-Laue-Str. 1, D-60438 Frankfurt, Germany\\
Tomsk State Pedagogical University, 634061 Tomsk, Russia\\
Lab.Theor.Cosmology,Tomsk State University of Control Systems and Radioelectronics (TUSUR), 634050 Tomsk, Russia\\
laurentis@th.physik.uni-frankfurt.de}

\maketitle
\begin{abstract}

We analyze a fully geometric approach to dark energy in the framework of $F(R,{\cal G})$ theories of gravity, where $R$ is the Ricci curvature scalar and ${\cal G}$ is the Gauss-Bonnet topological invariant. The latter invariant naturally exhausts, together with $R$, the whole curvature content related to  curvature invariants coming from the Riemann tensor. In particular, we study a class of $F(R, {\cal G})$ models with power law solutions and find that, depending on the value of the geometrical parameter, a shift in the anisotropy peaks position of the temperature power spectrum is produced, as well as an increasing in the matter power spectrum amplitude. This fact could be extremely relevant to fix the form of the $F(R, {\cal G})$  model. We also perform a MCMC analysis using both Cosmic Microwave Background data by the Planck (2015) release and the Joint Light-Curve Analysis of the SNLS-SDSS collaborative effort, combined with the current local measurements of the Hubble value, $H_0$, and galaxy data from the Sloan Digital Sky Survey (BOSS CMASS DR11). 
We show that such a model can describe the CMB data with slightly high $H_0$ values, and the prediction on the amplitude matter spectrum value is proved to be in accordance with the observed matter distribution of the universe. At the same time, the value constrained for the geometric parameter implies a density evolution of such a components that is growing with time. 

\end{abstract}

\ccode{PACS numbers: 04.50.-h, 04.20.Cv, 98.80.Jk}

\section{Introduction} 
A successful description of both the early and late time cosmic acceleration can be  addressed assuming a geometric approach in cosmology where the left hand side of the Einstein field equations is modified without adding further material contributions to the energy-momentum tensor~\cite{Sta80,Guth81,Sato81p1,Sato81p2,Kazanas80,PhysRepnostro,OdintsovPR,vasilis}. This `geometric picture'  can be realised  in several  ways considering   curvature or torsion invariants  into the effective gravitational action \cite{manos}. The general aim is to address  shortcomings of the standard cosmological model 
 \cite{ Kolb,Mukhanov81,Guth82,Hawking82,Sta82} both at early and late epochs under the standard of a comprehensive picture  that extend the good results of General Relativity. In this  sense, we deal with {Extended Theories of Gravity} \cite{PhysRepnostro}. 
 
In this approach the geometric invariant terms start contributing more significantly at a given epoch \cite{vasilis} and are directly related to important phenomena in the universe evolution, such as the process of structure formation and early and late-time cosmic acceleration. The theoretical predictions  of this geometric approach must  be confronted with observational data in order to verify their observational viability \cite{luongo1,luongo2,luongo3}. As a well known example, inflation generates  density perturbations with  a nearly scale invariant spectrum, a feature can be directly observed  measuring the Cosmic Microwave Background (CMB) anisotropies in temperature \cite{Planckdata,Planck2015,WMAP,Peiris03,Tegmark03,Tegmarkdata,2dF}. 
Currently, the viability of various inflationary models  \cite{Guth81,Sato81p1,Sato81p2, Kolb99, Linde83, Linde82, Albre82, Freese90, Polarski92, Linde94, PV99,FT02,Fe02,Paddy02,Sami02,SCQ02,TW05} have been investigated using these measurements.  

Both early and late time acceleration can be achieved  considering higher-order curvature terms in the action of gravitational interaction  \cite{PhysRepnostro,OdintsovPR,5,6,Mauro,faraoni, 9,10,libri,libroSV,libroSF}.  The  Starobinsky model \cite{Sta80}, for example, is an  inflationary scenario which is  realized considering  a $R^2$ contribution in  Ricci curvature scalar.  Other models realize inflation under a similar standard \cite{MO04, NOZ00, NO00, NO03, HHR01, ellis99, BB89, Maeda89}. From a fundamental physics point of view,  such  curvature invariants are derived as quantum corrections  from renormalisation of gravity in  curved spacetime \cite{Birrell}. Other  curvature invariants, as $R_{\mu\nu}R^{\mu\nu},R_{\mu\nu\sigma\rho}R^{\mu\nu\sigma\rho}$ have also been taken into account in literature   \cite{gorbu,ratbay,lorenzo1,lorenzo2,capcurve}, in particular 
the Gauss-Bonnet topological invariant  ${\cal G}$ which is related to the emergence of the trace anomaly in curved spacetime 
\cite{Birrell,Barth}.  In general, if both $R$ and ${\cal  G}$ are  present in the gravitational action, all the   curvature budget is considered, if we do not take into account further derivative terms like $\Box R$ and others \cite{nesseris}.   As shown in 
\cite{paolella}, the presence of  nonlinear ${\cal  G}$ terms in the action  gives rise to further inflationary episodes that can be connected to the observed large-scale structure process. In these models, one has an early  ${\cal G}$-dominated phase followed by the usual $R$-dominated phase. In general, the presence of Gauss-Bonnet topological invariant can solve some problems of $F(R)$ gravity, as discussed in~\cite{OdiGB, defelice, defelice1, fGBnoether, antonio1, topoquint, diego, myrz, DeLaurentis:2014oja}.  

{An important issue has to be addressed   after the gravitational wave event  GW170817, recently reported in \cite{waves}. As discussed in detail in \cite{ezquiaga}, many alternative theories of gravity can be  discarded considering  the upper bound on the wave propagation which is set to  be $|c_{g}/c - 1|\leq 5 \cdot 10^{-16}$ from the observations. In particular, $F(R)$ gravity remains a viable theory, while General Relativity, improved with $f(\phi){\cal G}$,  discussed in  \cite{sasaki}, seems to be excluded by the observations. There $f(\phi)$ is a function of a phantom  field and ${\cal G}$ is the Gauss-Bonnet invariant. The advantage of considering a Gauss-Bonnet non-minimal coupling relies on the fact  that an improved  phantom-quintessence phase of the  late universe  occurs thanks to   such a  term. The Gauss-Bonnet  curvature  becomes dominant and then the  phantom phase is a transient: this means that  the Big Rip singularity is avoided. This nice feature, in particular  the non-minimal coupling with a scalar field, seems in disagreement with recent measurement of GW170817 (see \cite{ezquiaga}). Despite of this fact, $F(R)$ remains  a viable theory and, better, $F(R,{\cal G})$ can be retained because it is a singularity free theory where the Gauss-Bonnet contribution enhances the reliable behavior of curvature quintessence \cite{capcurve} and inflation \cite{paolella}. This ghost-free behavior is one of the main motivation for the following study.}

In this paper, we test the observational viability of a class of $F(R,{\cal G})$ cosmologies and their power law solution for the scale factor derived in Ref. \cite{fGBnoether,topoquint}. 
 We perform a MCMC analysis using the current CMB data provided by the Planck Collaboration (2015)~\cite{Aghanim:2015xee}, along with type Ia supernovae observations from the Joint Light-Curve Analysis of the SNLS-SDSS collaborative effort~\cite{Betoule:2014frx}, current local measurements of the Hubble parameter, $H_0$~\cite{Riess}, and clustering data from the Sloan Digital Sky Survey (BOSS CMASS DR11)~\cite{Beutler:2013yhm}. We organize this paper as follows. Section~\ref{Sec:Model} reviews the basic features of $F(R,{\cal G})$ cosmology.  The method, the observational data sets,  and the priors used in the analysis are discussed in Section~\ref{Sec:Observational Analysis}. In Section~\ref{Sec:Results} we discuss our main results.  We summarize our main conclusions in Section~\ref{Sec:Conclusions}.

\section{The Gauss-Bonnet cosmology}
\label{Sec:Model}

Following the lines of Ref. \cite{fGBnoether} and adopting physical units such that $c= k_B=\hbar=1$, we consider the general action for the Gauss-Bonnet gravity
\begin{equation}
{\cal A}=\frac{1}{2\kappa}\int d^4x \sqrt{-g}\left[F(R,{\cal G})+{\cal L}_M\right]\,,
   \label{action}
\end{equation}
where $R$ is the Ricci scalar, $\cal G$ the  Gauss-Bonnet invariant and ${\cal L}_M$ the matter Lagrangian. Since we can define the Gauss-Bonnet invariant as
\begin{equation}
{\cal G}\equiv
R^2-4R_{\mu\nu}R^{\mu\nu}+R_{\mu\nu\rho\sigma}
R^{\mu\nu\rho\sigma}\,, 
   \label{GBinvariant}
\end{equation}
the action (\ref{action}) contains all the possible curvature invariants that can be derived starting from the Riemann tensor. Assuming a spatially flat  Friedmann-Robertson-Walker (FRW) metric, $ds^2=-dt^2+a^2(t)dx^idx_i$,  we can write the Friedmann equations as
\begin{subequations}  
\begin{eqnarray}
3f_{R}H^2&=&\kappa\rho^{(\mathrm{m})}+\frac{1}{2}(f_{R}R-F(R,{\cal G})-6H\dot{f_{R}}+\mathcal{G}f_{\mathcal{G}} \nonumber \\ & & -24H^3\dot{f_{\mathcal{G}}}),
\label{FRWa}
\end{eqnarray}
\begin{eqnarray}
2f_{R}\dot{H}&=&-\kappa\left(\rho^{(\mathrm{m})}+p^{(\mathrm{m})}\right)+H\dot{f_{R}}-\ddot{f_{R}}+4H^3\dot{f_{\mathcal{G}}}
\nonumber \\ & & -8H\dot{H}\dot{f_{\mathcal{G}}}-4H^2\ddot{f_{\mathcal{G}}}\label{FRWb} ,
\end{eqnarray}
\end{subequations}
where $a$ is the scale factor, $H=\dot{a}/a$ is the Hubble parameter, $\kappa =8\pi G$, $\rho^{(m)}$ and $p^{(m)}$ are the energy density and pressure of the clustered matter, respectively, and the overdot denotes a derivative with respect to the cosmic time $t$. In addition, here and henceforth, we use the notations
$f_{R}\equiv \frac{\partial F(R,\mathcal{G})}{\partial R}$ and $f_{\mathcal{G}}\equiv \frac{\partial F(R,\mathcal{G})}{\partial \mathcal{G}}$
for the partial derivatives with respect to $R$ and $\mathcal{G}$. The system of cosmological equations becomes self-consistent considering the definition of the Ricci curvature scalar and the Gauss-Bonnet invariant in terms of the scale factor $a$ and then the Hubble parameter $H$. As derived in  \cite{fGBnoether}, they are related to the Lagrange multipliers that have to be introduced in  the action \ref{action}. We have
\begin{eqnarray}
R &=& 6 \left[\frac{\ddot a}{a}+\left(\frac{\dot a}{a}\right)^2\right]= 6 \left(2H^{2}+\dot H \right)\,,
\label{eq:R} \\
{\cal G} &=&\frac{24 {\dot a}^2 {\ddot a}}{a^3}= 24H^{2} \left( H^{2}+\dot H \right)\,,  \label{eq:G}
\end{eqnarray}
defined for the signature $(-+++)$.
We can also rewrite the total energy density and pressure, $\rho_{(tot)}$ and $p_{(tot)}$, due to $R$ and Gauss-Bonnet contributions as \cite{DeLaurentis:2014oja}:
\begin{subequations}
\begin{eqnarray}
\rho_{(tot)} &= &\frac{1}{f_{R}}\left[\rho^{(m)}+\frac{1}{2\kappa}\left(R f_{R}-F(R,{\cal G})-6H\dot{f}_R+{\cal G}f_{\cal G}-24H^3\dot{f}_{\cal G}\right)\right]\,, \label{eq:rho-eff-1}
\\
\nonumber\\
p_{(tot)} &=& \frac{1}{f_{R}}\left\{p^{(m)}+\frac{1}{\kappa}\left[2H\dot{f}_R+\ddot{f}_{R}+8H^3\dot{f}_{\cal G}+8H\dot{H}\dot{f}_{\cal G}+4H^{2}\ddot{f}_{\cal G}-\frac{1}{2}\left(R f_{R}+{\cal G} f_{\cal G}-F(R,{\cal G}) \right)\right]\right\}
\,.
\label{eq:p-eff-1}
\end{eqnarray}
\end{subequations}
For cosmic acceleration, $\rho_{(tot)}+3 p_{(tot)} <0$, and assuming that all matter components have non-negative pressure, we can write the equation of state $w_{(\cal{GB})}=p_{(\cal{GB})}/\rho_{\cal{(GB})}$ from the geometry terms as 
%
 \begin{eqnarray}   \label{eos}
w_{(\cal{GB})}= \frac{{\cal G} f_{\cal G}
   +R f_{R}-F(R,{\cal G})-4 H \left[2 H {\ddot f}_{\cal G}+4 {\dot f}_{\cal G} \left(H^2+{\dot H}\right)+{\dot f}_{R}+ {\ddot f}_{R}/(2H)\right]}{F(R,{\cal G})+24 H^3 {\dot f}_{\cal G}-{\cal G} f_{\cal G} +6 H {\dot f}_{R}-R f_{R} }\,,
     \end{eqnarray} 
where the quintessence behavior is obtained for $-1\leq w_{(\cal{GB})}<0 $ while the phantom behaviour is achieved for $w_{(\cal{GB})}<-1$. 

Specifically, the Gauss-Bonnet term plays the role of a geometrical dark energy as in the case of $F(R)$ gravity\cite{capcurve} and then it  contributes to the effective cosmological density according to the formula
\begin{equation}
H(z) = H_0\left[\Omega_m a^{-3}+\Omega_r a^{-4}+\Omega_{\cal{(GB)}} a^{-3(1+  w_{(\cal{GB})})} \right]\,.
\end{equation}
This means that the effective value of the Gauss-Bonnet contribution can be ``measured" by evaluating the standard matter and radiation contributions at the various epochs, that is 
\begin{equation}
\Omega_{\cal{(GB)}}=1-\Omega_m - \Omega_r\,.
\end{equation}
It is important to stress that the coupling $f_R^{-1}$ does not affect the standard matter (and radiation) dynamics in Eqs. \ref{FRWa} and \ref{FRWb}, if we remain into the Jordan frame. In such a frame, the geodesic structure, tracked by matter particles and photons, is unaltered  and remains minimally coupled into the field equations. This means that only the geometrical part, i.e. the l.h.s. of field equations, is extended by assuming  $F(R,{\cal G})$ gravity. However, the situation completely change in the Einstein frame. Here, the result of the conformal transformation gives a non-minimally coupled  matter stress-energy tensor. This means that $\Omega_m$ has to be redefined while $\Omega_r$ results the same  because radiation is conformally invariant.   For a detailed discussion on this point,  see Ref. \cite{PhysRepnostro}.

\begin{figure}[!]
\centering
	\includegraphics[width=0.7\hsize]{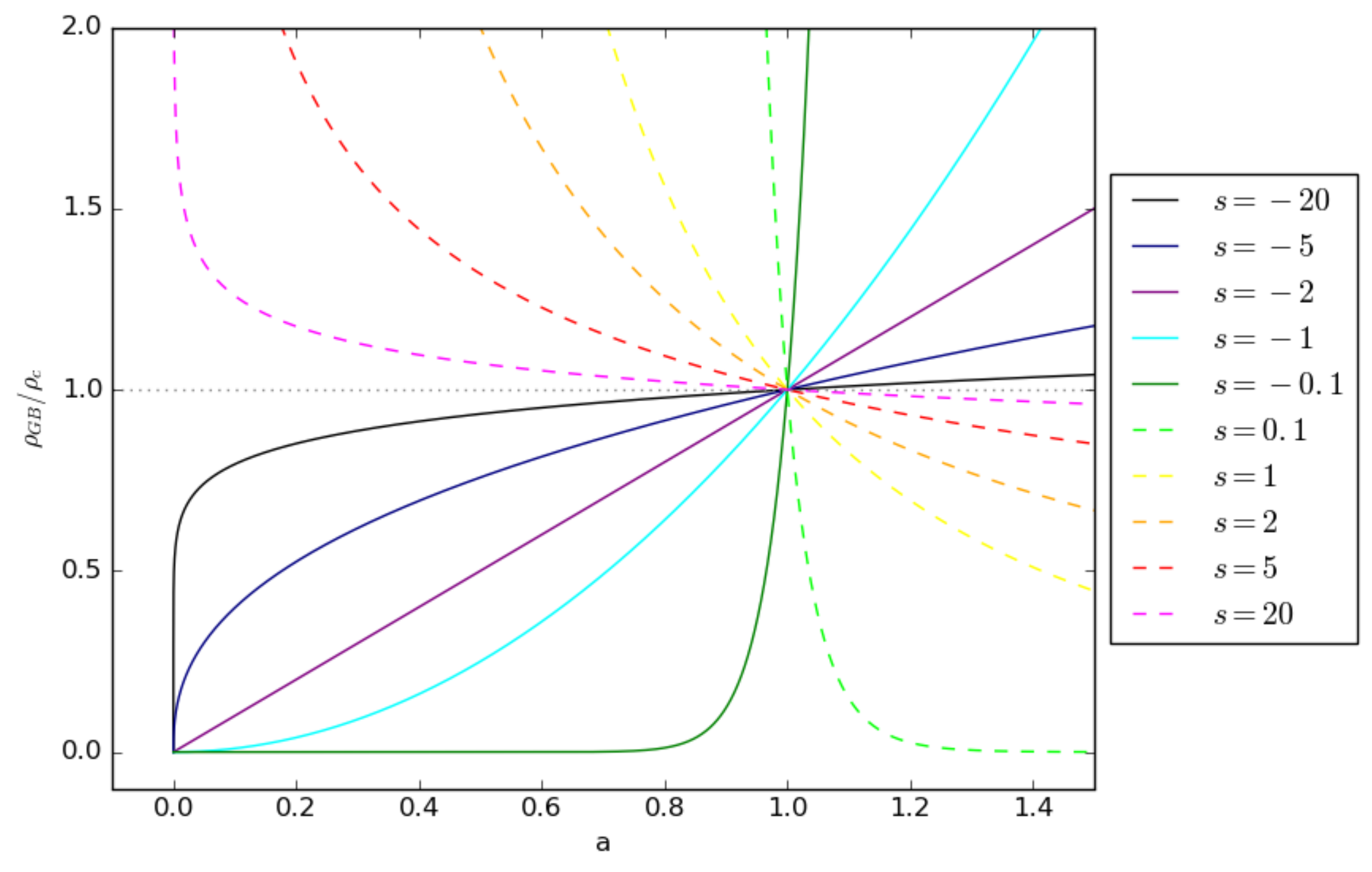}
	\includegraphics[width=0.7\hsize]{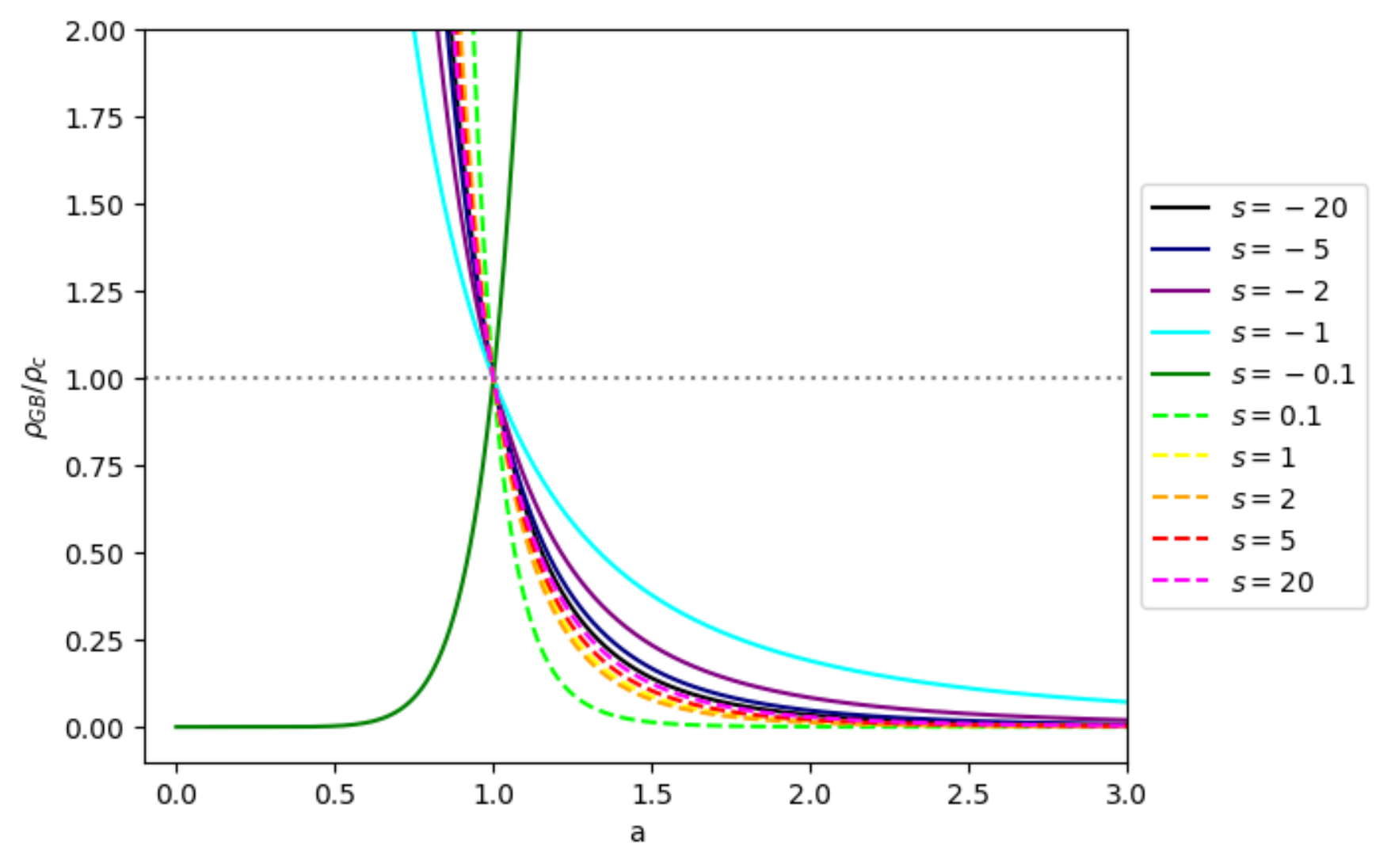}
	\caption{ {Evolution of the geometrical density $\rho_{\cal{(GB)}}$ for both solutions $n_1$ (top panel) and $n_2$ (bottom panel) with the scale factor $a$.}
	\label{fig:rho_1_2}}
\end{figure}

However, the form of $F(R,{\cal G})$ determines the evolution of $w_{(\cal{GB})}$. A viable $F(R,{\cal G})$ form can be achieved by considering the so called {\it Noether Symmetry Approach} \cite{fGBnoether}.
It can be shown that symmetries select the form of the function to be 
\begin{eqnarray}\label{FRG}
F({R,\cal G})= F_0 R^{n} {\cal G}^{1-n},
\end{eqnarray}
where $n$ is any real number and $F_0$ is a constant. For power law solutions of the form $a(t)=a_0t^s$ proposed in Ref.~\cite{topoquint} we obtain the relations
\begin{equation}
   n_1=\frac{1+s}{2} \quad {\mbox{and}} \quad n_2=\frac{1}{1+2s(s-1)}-2s\;,
\end{equation} 
such that the Eq. (\ref{eos}) can be written in terms of $n$ and $s$ as: 
%
\begin{figure}[!]
\centering
	\includegraphics[width=0.7\hsize]{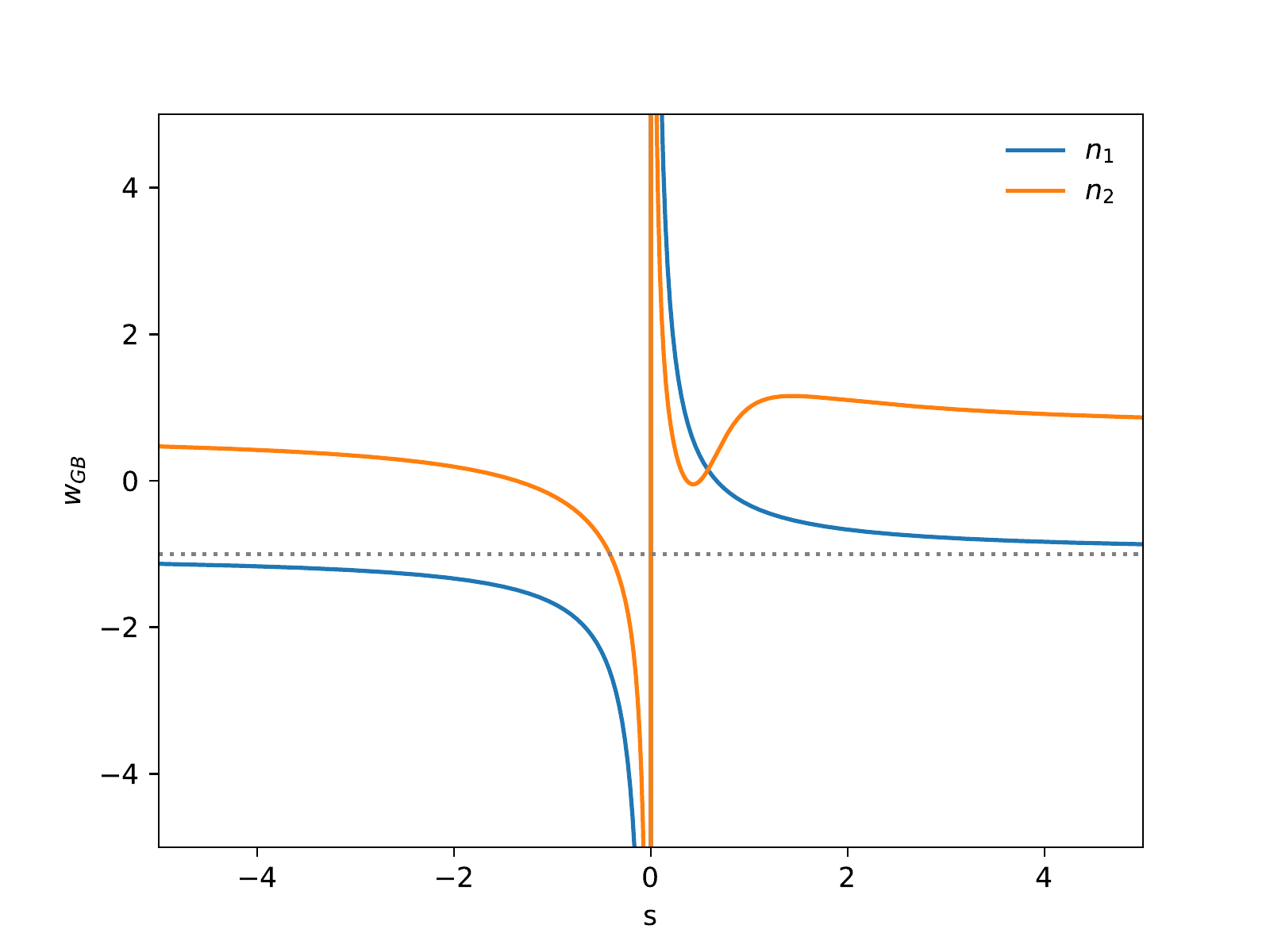}
	\caption{ {Behaviour of $w_{GB}$ with the geometrical parameter $s$ for both solutions $n_1$ and $n_2$. Notice that asymptotically $w_{GB}\rightarrow-1$ for the case $n_1$.}
	\label{fig:wGB_s}}
\end{figure}
%
\begin{equation}
\label{wzGB}
w_{(\cal GB)}=\frac{3 - 2( n + s )}{3 s}\;,
\end{equation}   
where $n$ can assume the values $n_1$ or $n_2$. 
In Fig.~\ref{fig:wGB_s} we show the behaviour of $w_{(\cal GB)}$ with the geometrical parameter $s$ for both relations $n_1$ and $n_2$. Note that asymptotically $w_{(\cal GB)}$ behaves like the $\Lambda$CDM model with ${\displaystyle w_{(\cal GB)}\rightarrow-1}$ for the case $n_1$. {{ Also, in Fig.~\ref{fig:rho_1_2} we show the evolution of the geometrical density, $\rho_{\cal{(GB)}}$, for both relations $n_1$ (top panel) and $n_2$ (bottom panel). In particular, we notice that for positive values of $s$, the geometrical density $\rho_{\cal{(GB)}}$ grows very rapidly  in the past and tends to zero today (for both $n_1$ and $n_2$). This would not allow a dynamic evolution of the universe dominated in the past by other components, i.e. matter and radiation, since this geometric component would be dominant. This also applies to positive values for the $n_1$ solution, while its negative values allow a density evolution increasing over time and dominant (only) today.}}
Since we aim to study viable values of the geometrical parameter, i.e. values that can well describe the observational data, we restrict our analysis to explore negative values of the parameter $s$ considering only the relation $n_1$. 

\begin{figure*}[t]
\centering
	\includegraphics[width=0.45\hsize]{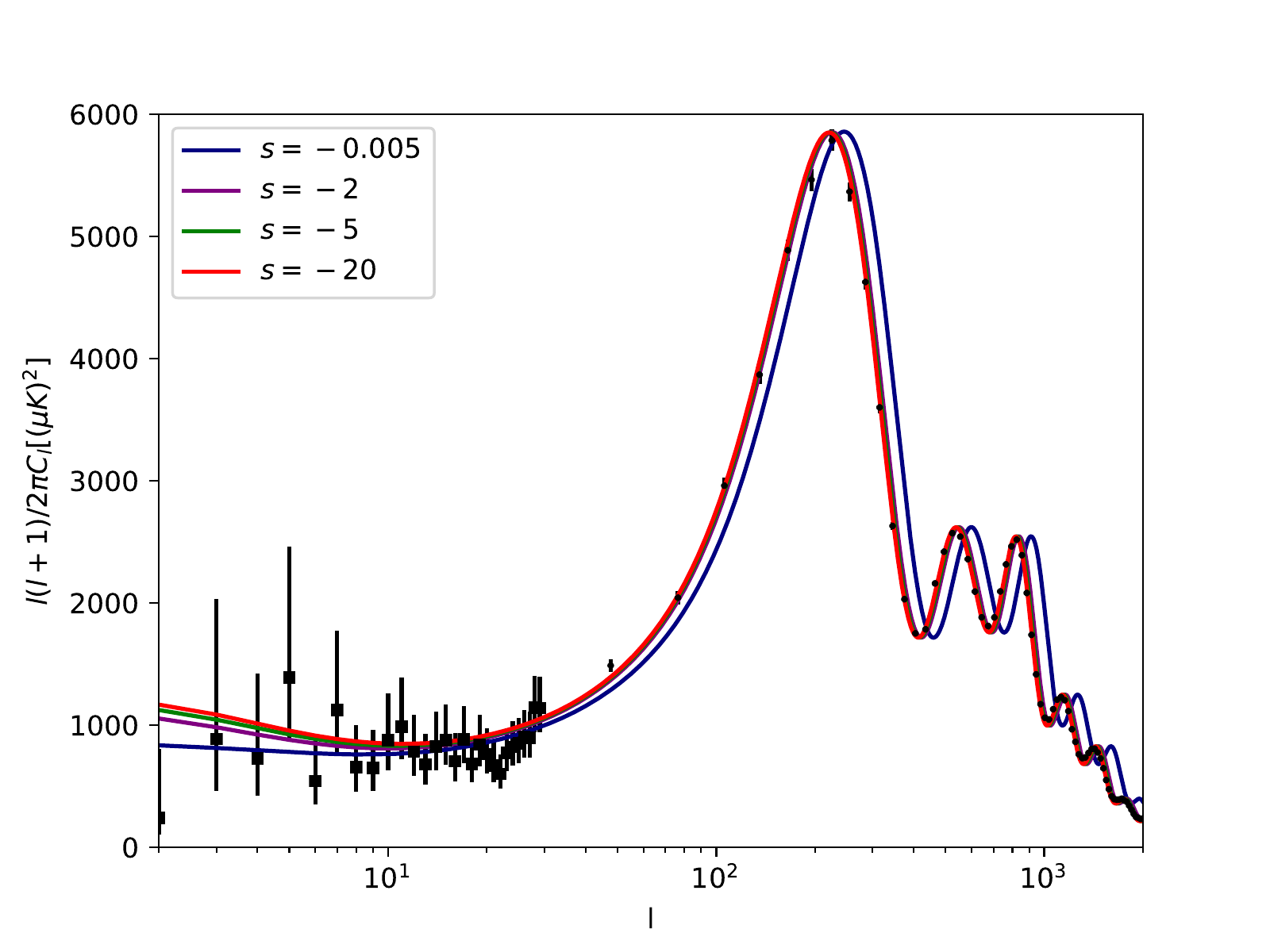}
	\includegraphics[width=0.45\hsize]{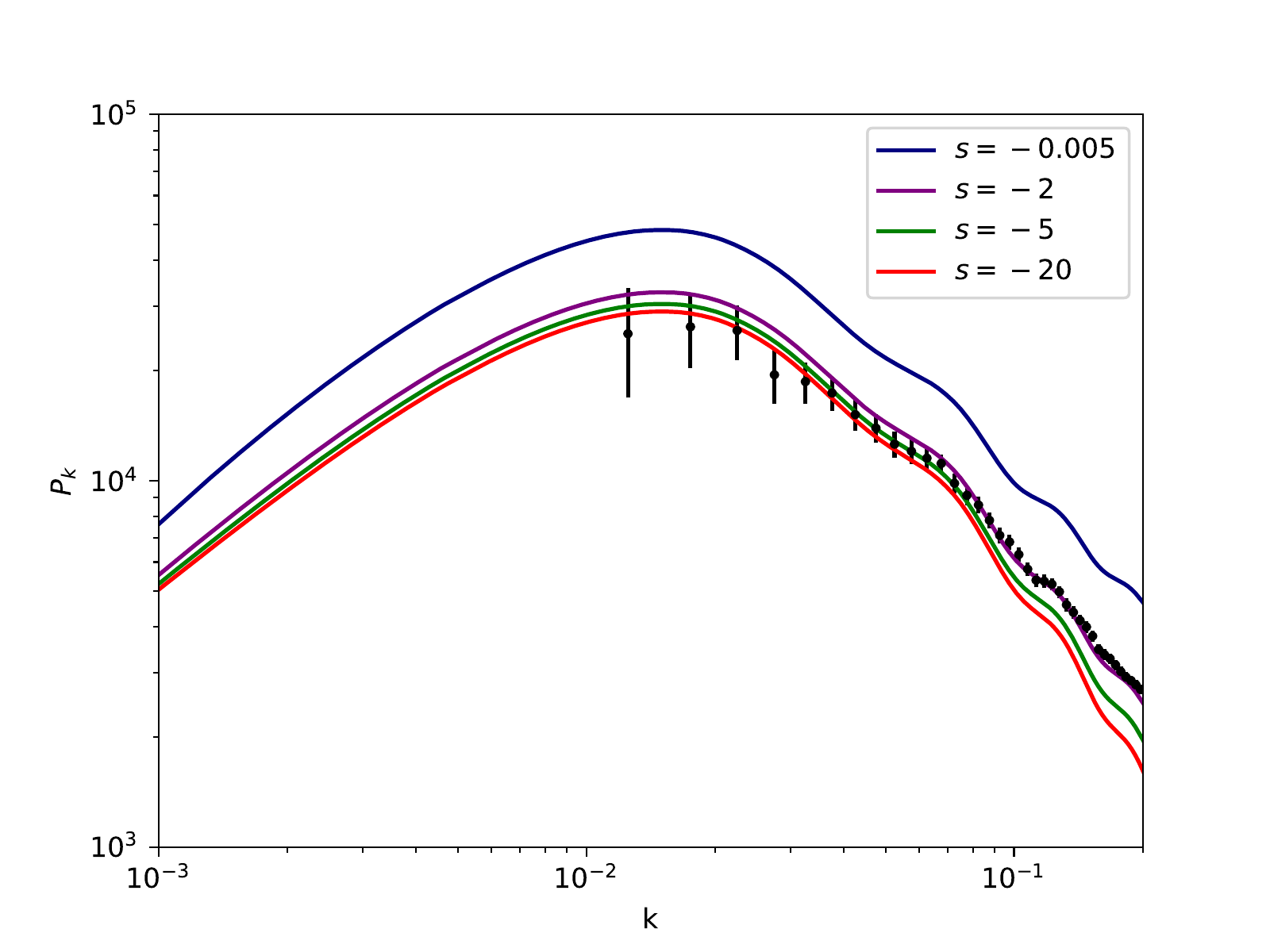}
	\caption{ {Anisotropy temperature (left panel) and matter power (right panel) spectra for several value of the geometrical parameter $s$. The data in the temperature power spectrum are from the binned Planck (2015) release, while for the matter power spectrum the data refers to the SDSS galaxy survey DR-11.}
	\label{fig:spectra}}
\end{figure*}

From the effect on the temperature power spectra, which is a slightly shift in the peaks position, and the influence in the amplitude in the matter power spectrum, as shown in Fig. \ref{fig:spectra}, we also note that for value of $s < -5$ the observational predictions remain almost unchanged (it is because $w_{(\cal GB)}$ is practically already $-1$), so we consider as a proper range for our analysis the flat prior $-20 < s <-0.005$.

%
%
\section{Method and Analysis}
\label{Sec:Observational Analysis}
In order to produce the observational prediction of the model, we use a modified version of the {\sc CAMB} code~\cite{camb}, where we introduce the geometrical parameter, $s$, as described in the previous section. We treat the geometrical component as a new cosmic fluid, considering the $w_{GB}$ contribution in both the background and in perturbative levels.
We compare the model predictions with the data by a Monte Carlo Markov chain analysis, using the available package {\sc CosmoMC}~\cite{cosmomc}.
In our analysis, in addition to the geometrical parameter we also vary the usual set of cosmological variables, namely the baryon and the cold dark matter density, the ratio between the sound horizon, the angular diameter distance at decoupling, the optical depth, the primordial scalar amplitude and spectral index :
$\left \{\Omega_bh^2~,~\Omega_ch^2~,~\theta~,~\tau,~A_s,~n_s \right \}$.
We consider purely adiabatic initial conditions, fix the sum of neutrino masses to $0.06~eV$ and the universe curvature to zero, and also vary the nuisance foregrounds parameters~\cite{Aghanim:2015xee}.
The large flat priors we used on the cosmological and geometrical parameters are shown in Table \ref{tab_priors}. 

We use the CMB data set from the latest Planck (2015) Collaboration release~\cite{Aghanim:2015xee}, considering the high multipoles Planck temperature data  from the 100-,143-, and 217-GHz half-mission T maps, and  the low multipoles data by the joint TT, EE, BB and TE likelihood, where EE and BB are the E- and B-mode CMB polarization power spectrum and TE is the cross-correlation temperature-polarization (hereafter ``PLC2"). 
We also combine the CMB data  with an extended background data sets, composed of Supernovae Type (SNe) Ia, Hubble constant local measurement and galaxy data, i.e.:
\begin{itemize}
\item {for the SNe data, we use the ``Joint  Light-curve  Analysis"  (JLA) sample~\cite{Betoule:2014frx};}
\item {for the $H_0$ measurement we use the Riess {\it{et al.}} (2016) results on the local expansion rate~\cite{Riess}, $H_0 = 73.24 \pm 1.74$ $\rm{km.s^{-1}.Mpc^{-1}}$ (68\% C.L.), based on direct measurements made with the Hubble Space Telescope (HST). This measurement is used as an external Gaussian prior;}
\item {for the  galaxy data, {{we use the full matter power spectrum by  measurements from the}}  Baryon Oscillation Spectroscopic  Survey  (BOSS)  CMASS  Data  Release-11 sample  (covering  the  redshift  range $z= 0.43-0.7$)  of the Sloan Digital Sky Survey experiment (SDSS)~\cite{Beutler:2013yhm}, publicly available in the SDSS Collaboration website (www.sdss3.org).}
\end{itemize}
Th extended data set (Ext) used in our analysis comprises PLC2, HST, JLA and SDSS data.
%

%
\begin{table}
\centering
\tbl{Priors on the model parameters.}
{\begin{tabular}{|c|c|}
\hline 
Parameter & Prior Ranges \\ 
\hline
$\Omega_{b}h^{2}$ & $[0.005 : 0.1]$ \\ 
 
$\Omega_{c}h^{2}$ & $0.001 : 0.99]$ \\ 

$\theta$ & $[0.5 : 10.0]$ \\ 
 
$\tau$ & $[0.01 : 0.8]$ \\ 
 
$n_{s}$ & $[0.8 : 1.2]$ \\ 
 
$\log{10^{10}}A_{s}$ & $[2.0 : 4.0]$ \\ 
 
$s$ & $[-20 : -0.005]$ \\ 
\hline 
\end{tabular}\label{tab_priors}}
\end{table} 

\begin{table}
\centering
\tbl{$68\%$ confidence limits and best fit values for the cosmological and geometrical parameters. The  first  columns-block show the constraints for the $w_{\cal GB}$ model using PLC2 data, while the second  columns-block  refers to constraints using the Ext data set, i. e. joint PLC2+HST+JLA+SDSS data. 
The table is divided into two sections: the upper section are the primary parameters, while in the lower part are the derived ones.
The last line stands for $\Delta\chi^2_{best}$ values, which refers to the difference with respect to the $\Lambda$CDM model using the same data set. Note that the upper limit of the geometrical parameter $s$ refers at $95\%$ C.L..}
{\begin{tabular}{|c|c|c|}
\hline
\hline
{Parameter}& {PLC2} & {Ext} 
\\
\hline
Primary & & \\
{$100\,\Omega_b h^2$} 
& $2.224 \pm 0.024$
& $2.228 \pm 0.022$
\\
{$\Omega_{c} h^2$} 
& $0.1196 \pm 0.0022$
& $0.1193 \pm 0.0018$
\\
{$\theta$}
& $1.04090 \pm 0.00048$
& $1.04098 \pm 0.00044$
\\
{$\tau$}
& $0.077 \pm 0.020$
& $0.079 \pm 0.019$
\\
{$\ln 10^{10}A_s$  }  
& $3.088 \pm 0.038$
& $3.091 \pm 0.037$
\\
{$n_s$}
& $0.9657 \pm 0.0064$
& $0.9670 \pm 0.0055$
\\
{$s$}
& $<-0.005$
& $-10.94 \pm 4.35$
\\
\hline
\hline
Derived & & \\
{$H_{0}$}
& $73.26 \pm 6.91$
& $69.74 \pm 0.99$
\\
{$\Omega_{m}$}
& $0.272 \pm 0.043$
& $0.293 \pm 0.010$
\\
{$\Omega_{\cal GB}$}
& $0.728 \pm 0.043$
& $0.707 \pm 0.010$
\\
{$w_{(\cal GB)}$}
& $> -134$
& $-1.06 \pm 0.02$
\\
\hline
\hline
{$\Delta \chi^2_{\rm best}$}  
& $-1.4$
& $3.6$
\\
\hline
\end{tabular} \label{tab:Tabel_results_1}}
\end{table} 

%
\begin{figure*}[]
   \includegraphics[width=1.\hsize]{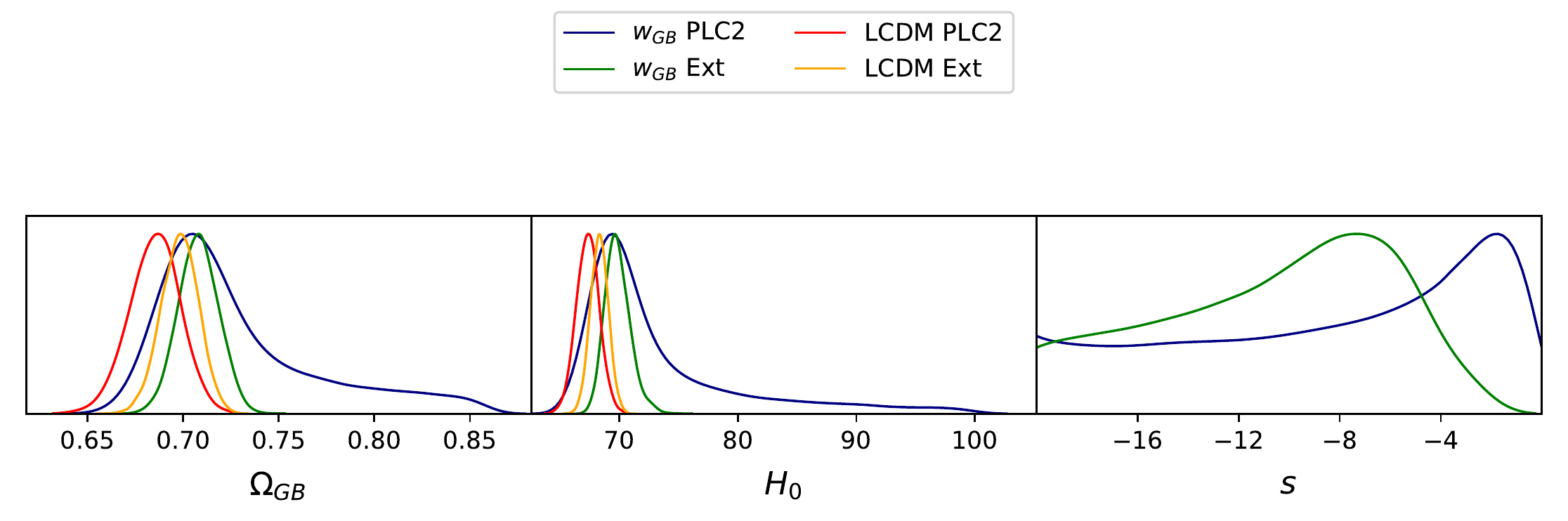}
	\caption{ {Probability distribution for the cosmological parameters of the $w_{(\cal GB)}$ model using the PLC2 (green line) and Ext (blue line) data set in comparison with the $\Lambda$CDM model using PLC2 (red line) and Ext (black line) data set. The total density of the geometry contribution, $\Omega_{GB}$, is compared with the dark energy density, $\Omega_\Lambda$, of the standard cosmological model.}
	\label{fig:analysis_parameters1}}
\end{figure*}

%
\begin{figure*}[]
\centering
    \includegraphics[width=0.45\hsize]{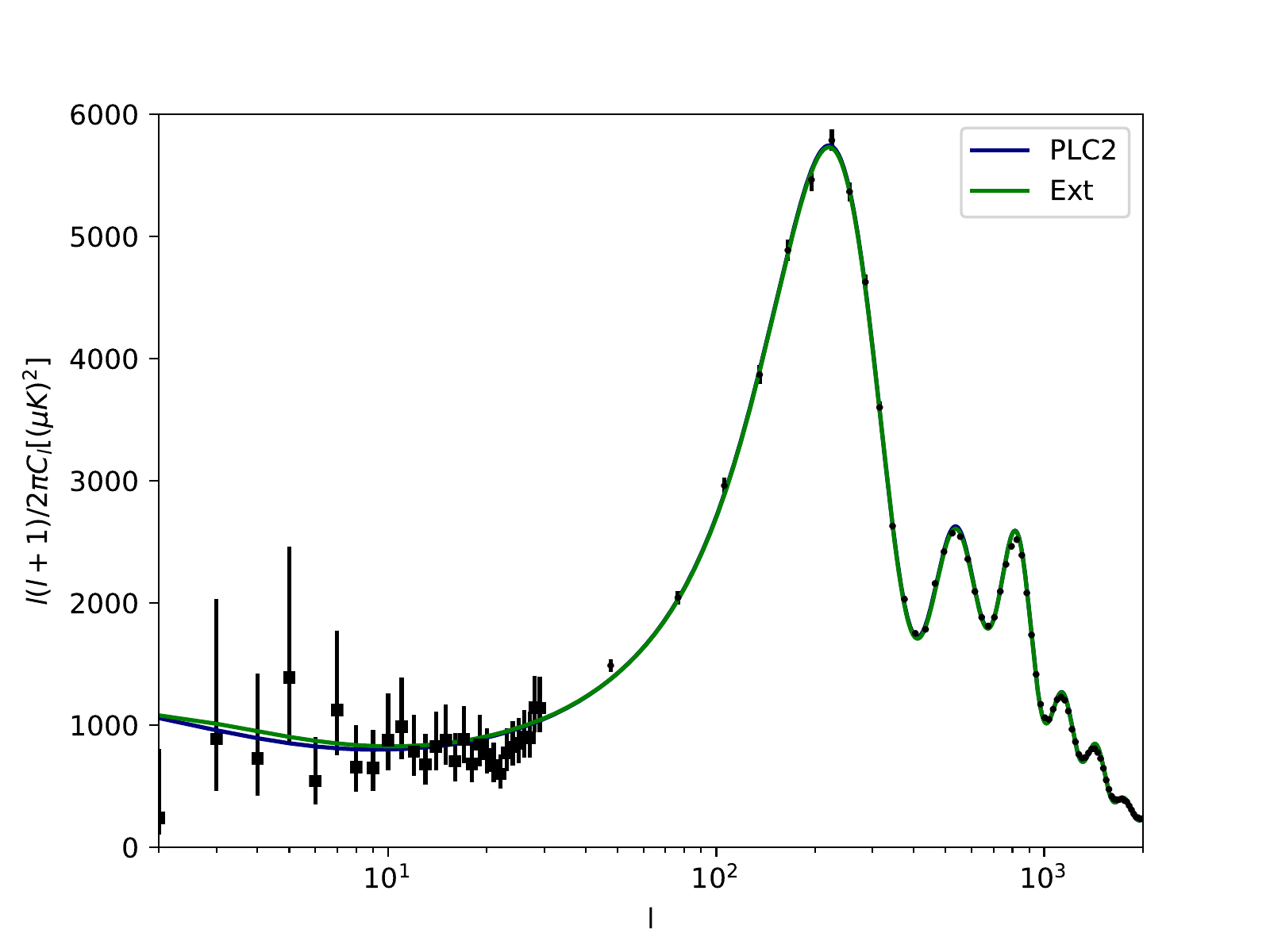}
	\includegraphics[width=0.45\hsize]{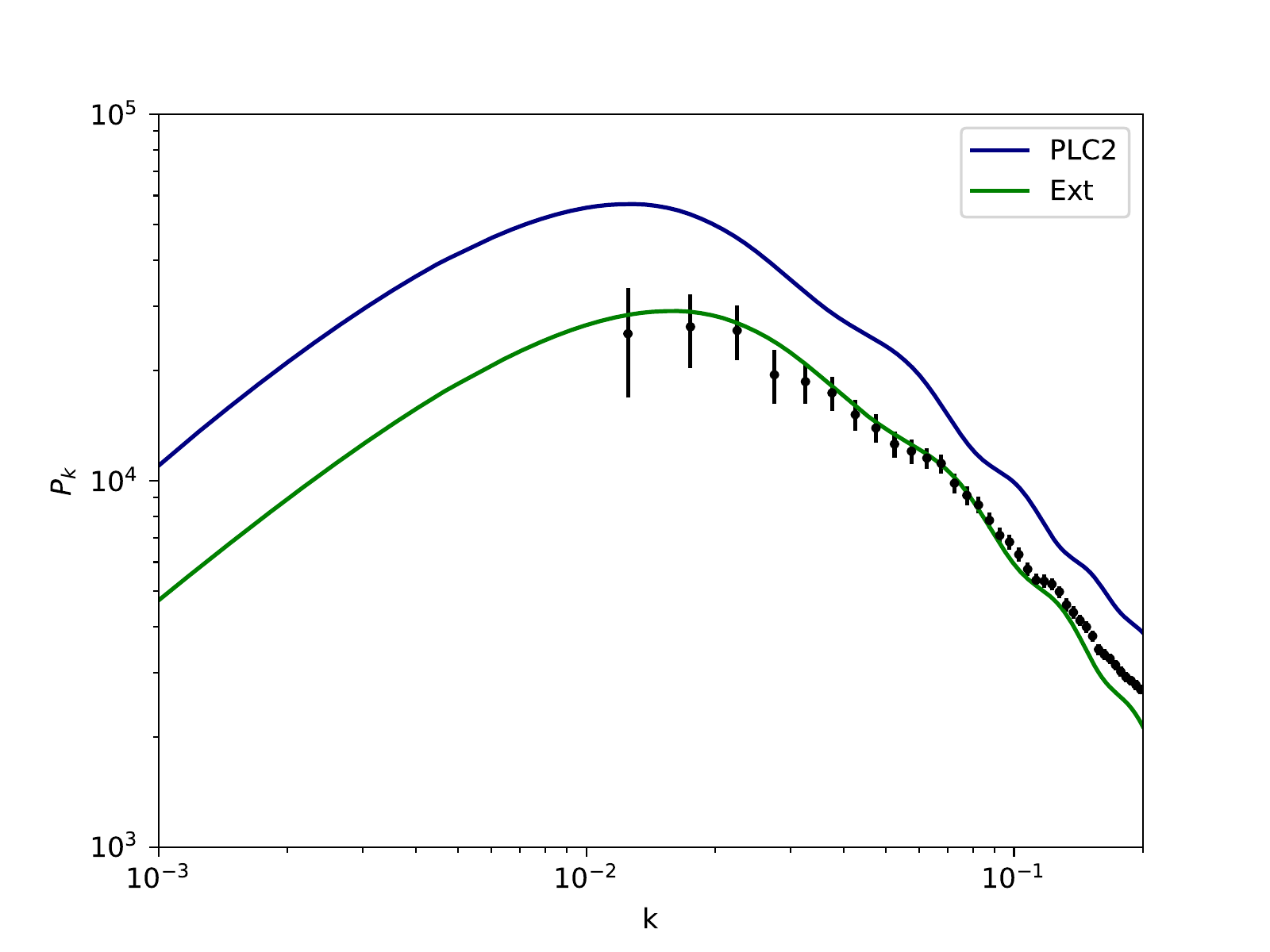}
	\caption{ { Anisotropy temperature (left panel) and matter power (right panel) spectra for the $w_{\cal GB}$ model best fit values using the PLC2 and Ext data. In the temperature power spectrum the data are from the binned Planck (2015) release, while for the matter power spectrum the data refers to the SDSS galaxy survey DR11.}
	\label{fig:analysis_spectrum1}}
\end{figure*}

\section{Results}
\label{Sec:Results}

The main results of our analysis are summarized in Tab.\ref{tab:Tabel_results_1}, which displays the bounds on the cosmological and geometrical parameters. We also show in Fig. \ref{fig:analysis_parameters1} the posterior probability distributions of $s$ and the derived parameters $H_0$ and $\Omega_{GB}$, namely the Hubble constant and the density parameter of the geometry contribution. For comparison we also show the constraints on the dark energy density parameter, $\Omega_{\Lambda}$, of the standard $\Lambda$CDM model.

From the first column of Tab.~\ref{tab:Tabel_results_1} one can see that while the constraints on the primary parameters agree with the $\Lambda$CDM results (see, e.g.,~\cite{Ade:2015xua}) we find a larger value for the derived Hubble parameter, which disagrees at $1\sigma$ with the $\Lambda$CDM prediction (see also the green line of Fig.~\ref{fig:analysis_parameters1}). Also, the contribution of the $\Omega_{GB}$ parameter is higher with respect to the dark energy density of the $\Lambda$CDM model.
Such differences can be explained if one considers that the effect of the geometrical contribution on the temperature power spectra is a shift in the peaks position and in the matter power spectrum amplitude (see left and right panel on Fig.\ref{fig:spectra}). Indeed, these signatures are related, in the standard cosmological model, with the $\Omega_\Lambda$ and the $H_0$ values. 
It means that, since the geometric parameter in the $w_{\cal GB}$ context produce a shift in the anisotropy peaks like the one produced by $\Omega_\Lambda$ and the $H_0$ values in the $\Lambda$CDM context, these latter must assume higher values to compensate such a shift. As one may  see the posterior distribution of the $s$ parameter is not Gaussian. This parameter is very weakly constrained using only the PLC2 data, while the constraint improves when one considers the Ext data set (as well as the constraints on $H_0$ and $\Omega_{GB}$). {{We also note that the $w_{(\cal GB)}$ parameter falls slightly in the phantom regime for the Ext data set, and it shows strong phantom behaviour using only the PLC2 data.}}

The best fit of the $w_{\cal GB}$ model using the PLC2 (blue) and the Ext (green) data is shown in Fig.\ref{fig:analysis_spectrum1}. We note that both of them are in accordance with the Planck (2015) data for the temperature anisotropy power spectrum (left panel), while the Ext data set better describes the SDSS-DR11 data in the matter power spectrum (right panel). Finally, the $\Delta\chi^2$ with respect to the $\Lambda$CDM model are reported in the last line of Tab.~II. {{Even it is not possible to extract significant statistical information by this simple comparison of $\chi^2$ values \footnote {We refer the reader to Ref. \cite{Trotta, Trotta:2017wnx} for a discussion on proper model comparison.}, we can see that the minimum likelihood of the best fit of the analysed  $w_{\cal GB}$ model is close to the standard model one.}}

It is worth noting that the constraint on the $s$ parameter is not tight nor Gaussian, as has been mentioned earlier. 
It depends upon the limit ${\displaystyle w_{(\cal GB)}\rightarrow-1}$ (see Fig.~\ref{fig:wGB_s}), so that for values of $s <-5$ there are few changes in the observational predictions, i.e., all the values of $s$ produce almost the same observable. Moreover, the constrained best fit value corresponds to a density evolution $\frac{\rho_{(\cal GB)}(z)}{\rho_{c,0}}=a^{-3(1+w_{GB})} \sim a^{0.3}$, which means a density that is growing with time.
\section{Conclusions}
\label{Sec:Conclusions}

$F(R,{\cal G})$ theories of gravity take into account all the curvature budget coming from the Riemann tensor by combining the Ricci curvature scalar and the Gauss-Bonnet topological invariant. This class of models can be related to  quantum field theories on curved spacetimes  \cite{Birrell,Barth} and  has been largely investigated  as an generalisation of the $F(R)$ gravity at the cosmological level \cite{paolella}. 

In this paper, we have examined the observational viability of power law solutions~\cite{topoquint} for a class of $F(R,{\cal G})$ models derived from   Noether's symmetries whose form of the $F(R,{\cal G})$ function is given by  $F(R,{\cal G})=F_0 R^{n} {\cal G}^{1-n}$ \cite{fGBnoether}. We have shown that for a subsample of model parameters viable accelerating solutions can be achieved from geometrical terms, described in the form of a modified equation of state parameter $w_{(\cal GB)}$. In some sense, this is an extension of the approach already developed for $F(R)$ gravity \cite{capcurve}.

Confronting these models with data, we have shown that they predict a shift in the position of the peaks of the CMB temperature power spectrum and an increasing in the matter power spectrum amplitude with respect to the standard model. We have tested the observational viability of this approach by comparing its theoretical predictions with joint data of CMB, SNe Ia, local $H_0$ measurement and the matter power spectrum. We have found that both the anisotropy temperature data and the matter distribution are well fitted, with the price of an high total density of the geometry contribution, $\Omega_{GB}$, and $H_0$ values.  
We conclude that this geometrical description, assuming  power law solutions, can describe the current available observational data without further dark energy contributions. Remarkable, its density evolution is a function that is growing with time. 

{An important remark is necessary at this point. Solutions $n_1$ and $n_2$ have a power-law behavior as determined by the Noether symmetries. This  assumption could seem too simple in order to address the whole cosmological evolution starting from inflation, evolving into radiation/matter dominated eras and, finally, ending up into the  dark energy behavior. However, as discussed in detail in Ref. \cite{antonio},  it is possible to show that solutions of the type
\begin{equation}
a = a_0(t - t_0)^{\frac{2n}{3(1+w)}} , 
\end{equation}
arises as transient phases in any extended theory  like, for example,  $F(R)$ gravity. Here, as above, 
 $w$ is a  barotropic index. These transient  phases can  evolve into  accelerated behaviors  representing  attractors  for the dynamical system describing the related cosmology \cite{sante}. In any case, the same exact  solution,  matching together the  sequence  of  inflation, radiation, matter and dark energy is unrealistic to be found out in any   power law theory of gravity. Nevertheless, considering transient behaviors like in the case of $n_1$ and $n_2$ can be highly indicative to figure out the evolution of a single cosmological era.
In a forthcoming  paper, we will generalize the present results  performing a dynamical system analysis
for  generic  $F(R, {\cal G})$ models ~\cite{Simony}.}

\section*{Acknowledgements}

The authors acknowledge Fernando V. Roig for useful conversations and also Anthony Lewis for the use of the CosmoMC code. M. Benetti is supported by the Funda\c{c}\~{a}o Carlos Chagas Filho de Amparo \`{a} Pesquisa do Estado do Rio de Janeiro (FAPERJ - fellowship {\textit{Nota 10}}). 
S. Santos da Costa acknowledges financial support from Coordena\c{c}\~{a}o de Aperfei\c{c}oamento de Pessoal de N\'ivel Superior (CAPES).
S. Capozziello is supported in part by the INFN sezione di Napoli, iniziative specifiche TEONGRAV and QGSKY. J. Alcaniz acknowledges support from CNPq (Grants no. 310790/2014-0 and 400471/2014-0) and FAPERJ (grant no. 204282). 
{{M. De Laurentis  is supported by ERC Synergy Grant Imaging the Event Horizon of Black Holes awarded by the ERC in 2013 (Grant No. 610058)}}. This article is based upon work from COST Action CA15117  ``Cosmology and Astrophysics Network for Theoretical Advances and Training Actions" (CANTATA), supported by COST (European Cooperation in Science and Technology).


\begin{thebibliography}{99}

\bibitem{Sta80}
A.~A.~Starobinsky,
{\it Phys.\ Lett.\ B} {\bf 91}, 99 (1980).
\bibitem{Guth81}
A.~H.~Guth,
{\it Phys.\ Rev.\ D} {\bf 23}, 347 (1981).
\bibitem{Sato81p1}
K.~Sato, {\it Mon. Not. R. Astron. Soc.} {\bf 195}, 467 (1981).
\bibitem{Sato81p2}
K.~Sato, {\it Phys. Lett. B} {\bf 99}, 66 (1981).
\bibitem{Kazanas80}
D.~Kazanas,
{\it Astrophys.\ J.} {\bf 241} L59 (1980).
\bibitem{PhysRepnostro}S. Capozziello, M. De Laurentis, {\it Phys. Rept.} {\bf 509}, 167 (2011).
\bibitem{OdintsovPR} S. Nojiri, S. D. Odintsov, {\it Phys. Rept.} {\bf 505}, 59 (2011).
%
\bibitem{vasilis}
S. Nojiri, S.D. Odintsov, V.K. Oikonomou, {\it Phys. Rept.} {\bf 692}, 1  (2017).

\bibitem{manos}
  Y.~F.~Cai, S.~Capozziello, M.~De Laurentis and E.~N.~Saridakis,
  Rept.\ Prog.\ Phys.\  {\bf 79}, 106901  (2016).

%
\bibitem{Kolb}
E.~W.~Kolb and Turner,
{\em The Early Universe},
Addison-Wesley, Redwood City  (1990).
\bibitem{Mukhanov81}
V.~F.~Mukhanov and G.~V.~Chibisov,
{\it JETP Lett.}  {\bf 33}  532 (1981);{\it Pisma Zh.\ Eksp.\ Teor.\ Fiz.} {\bf 33}, 549 (1981).
\bibitem{Guth82}
A.~H.~Guth and S.~Y.~Pi,
{\it Phys.\ Rev.\ Lett.} {\bf 49}, 1110 (1982).
\bibitem{Hawking82}
S.~W.~Hawking,
{\it Phys.\ Lett.\ B} {\bf 115}, 295 (1982).
\bibitem{Sta82}
A.~A.~Starobinsky,
{\it Phys.\ Lett.\ B} {\bf 117}, 175 (1982).
%
\bibitem{luongo1}
  A.~de la Cruz-Dombriz, P.~K.~S.~Dunsby, O.~Luongo and L.~Reverberi,
  JCAP {\bf 1612} (2016) no.12,  042
\bibitem{luongo2}
  P.~K.~S.~Dunsby, O.~Luongo and L.~Reverberi,
  Phys.\ Rev.\ D {\bf 94},  083525 (2016).
%
\bibitem{luongo3}
  P.~K.~S.~Dunsby and O.~Luongo,
  Int.\ J.\ Geom.\ Meth.\ Mod.\ Phys.\  {\bf 13},  1630002 (2016).
  %
\bibitem{Planckdata}Planck Collaboration, {\it Astron. Astrophys.} {\bf 571}, A22 (2014). 
\bibitem{Planck2015}Planck Collaboration, arXiv:1502.02114 [astro-ph.CO] (2015); 	arXiv:1502.01590 [astro-ph.CO] (2015); 	arXiv:1502.01589 [astro-ph.CO] (2015).
\bibitem{WMAP}
D.~N.~Spergel {\it et al.},
{\it Astrophys.\ J.\ Suppl.} {\bf 148}, 175 (2003).
\bibitem{Peiris03}
H.~V.~Peiris {\it et al.},
{\it Astrophys.\ J.\ Suppl.}  {\bf 148}, 213 (2003).
\bibitem{Tegmark03}
M.~Tegmark {\it et al.}  [SDSS Collaboration],
{\it Phys.\ Rev.\ D} {\bf 69}, 103501 (2004).
\bibitem{Tegmarkdata}
M.~Tegmark {\it et al.}  [SDSS Collaboration],
Astrophys.\ J.\  {\bf 606}, 702 (2004).
\bibitem{2dF}
W.~J.~Percival {\it et al.},
{\it Mon. Not. Roy. Astron. Soc.} {\bf 327}, 1297 (2001).
\bibitem{Kolb99}
E.~W.~Kolb, {\it Pritzker Symposium and Workshop on the Status of Inflationary Cosmology}, arXiv:hep-ph/9910311 (1999).
\bibitem{Linde83}
A.~D.~Linde,
{\it Phys.\ Lett.\ B} {\bf 129} 177 (1983).
\bibitem{Linde82}
A.~D.~Linde,
{Phys.\ Lett.\ B} {\bf 108}, 389 (1982).
\bibitem{Albre82}
A.~Albrecht and P.~Steinhardt,
{\it Phys.~Rev.~Lett.} {\bf 48}, 1220 (1982).
\bibitem{Freese90}
K.~Freese, J.~A.~Frieman and A.~V.~Olinto,
{\it Phys.\ Rev.\ Lett.} {\bf 65}, 3233 (1990).
\bibitem{Polarski92}
D.~Polarski and A.~A.~Starobinsky,
{\it Nucl.\ Phys.\ B} {\bf 385}, 623 (1992).
\bibitem{Linde94}
A.~D.~Linde,
{\it Phys.\ Rev.\ D} {\bf 49}, 748 (1994).
\bibitem{PV99}
P.~J.~E.~Peebles and A.~Vilenkin,
{\it Phys.\ Rev.\ D} {\bf 59}, 063505 (1999).
\bibitem{FT02}
M.~Fairbairn and M.~H.~G.~Tytgat,
{\it Phys.\ Lett.\ B} {\bf 546}, 1 (2002).
\bibitem{Fe02}
A.~Feinstein,
{\it Phys.\ Rev.\ D} {\bf 66}, 063511 (2002).
\bibitem{Paddy02}
T.~Padmanabhan,
{\it Phys.\ Rev.\ D} {\bf 66}, 021301 (2002).
\bibitem{Sami02}
M.~Sami,
{\it Mod.\ Phys.\ Lett.\ A} {\bf 18}, 691 (2003).
\bibitem{SCQ02}
M.~Sami, P.~Chingangbam and T.~Qureshi,
{\it Phys.\ Rev.\ D} {\bf 66}, 043530 (2002).
\bibitem{TW05}
S.~Thomas and J.~Ward,
{\it Phys.\ Rev.\ D} {\bf 72}, 083519 (2005).
\bibitem{5}S. Nojiri, S. D. Odintsov,  {\it eConf. C} {\bf 0602061}, 06 (2006).
\bibitem{6}S. Nojiri, S. D. Odintsov, {\it Int.\ J.\ Geom.\ Meth.\ Mod.\ Phys.} {\bf 4}, 115 (2007).
\bibitem{Mauro}S. Capozziello, M. Francaviglia, {\it Gen. Rel. Grav.} {\bf 40}, 357 (2008).
\bibitem{faraoni} S. Capozziello, M. De Laurentis, V. Faraoni, {\it The Open Astr. Jour} , {\bf 2}, 1874 (2009).
\bibitem{9} A. de la Cruz-Dombriz and D. Saez-Gomez, {\it Entropy} {\bf 14}, 1717 (2012).
\bibitem{10}G.~J.~Olmo, {\it Int.\ J.\ Mod.\ Phys.\ D} {\bf 20}, 413  (2011).
\bibitem{libri}
F. S. N. Lobo, {\it Dark Energy-Current Advances and Ideas}, 173-204 (2009), Research Signpost, ISBN 978-81-308-0341-8arXiv:0807.1640 [gr-qc].
\bibitem{libroSV}S. Capozziello and V. Faraoni,  {\it Beyond Einstein gravity: A Survey of gravitational theories for cosmology and astrophysics.} Fundamental Theories of Physics. 170. Springer. (2010),  ISBN 978-94-007-0164-9.
\bibitem{libroSF}S. Capozziello and M. De Laurentis, {\it Invariance Principles and Extended Gravity:
Theories and Probes}, Nova Science Publishers, Inc. (2010) ISBN: 978-1-61668-500-3.
\bibitem{MO04}
K.~i.~Maeda and N.~Ohta,
{\it Phys.\ Lett.\ B} {\bf 597}, 400 (2004).
\bibitem{NO00}
S.~Nojiri and S.~D.~Odintsov,
{\it Phys.\ Lett.\ B} {\bf 484}, 119 (2000).
\bibitem{NOZ00}
S.~Nojiri, S.~D.~Odintsov and S.~Zerbini,
{\it Phys.\ Rev.\ D} {\bf 62}, 064006 (2000).
\bibitem{NO03}
S.~Nojiri and S.~D.~Odintsov,
{\it Phys.\ Rev.\ D} {\bf 68}, 123512 (2003).
\bibitem{HHR01}
S.~W.~Hawking, T.~Hertog and H.~S.~Reall,
{\it Phys.\ Rev.\ D} {\bf 63}, 083504 (2001).
\bibitem{ellis99}
J.~R.~Ellis, N.~Kaloper, K.~A.~Olive and J.~Yokoyama,
{\it Phys.\ Rev.\ D} {\bf 59}, 103503 (1999).
\bibitem{BB89}
M.~C.~Bento and O.~Bertolami,
{\it Phys.\ Lett.\ B} {\bf 228}, 348 (1989).
\bibitem{Maeda89}
K.~I.~Maeda,
{\it Phys.\ Rev.\ D} {\bf 39}, 3159 (1989).
\bibitem{Birrell}  N.D. Birrell,  P.C.W. Davies, {\it
Quantum Fields in Curved Space}, Cambridge University Press,
Cambridge, UK (1982).
%
\bibitem{gorbu} D. Gorbunov and A. Tokareva, {\it Phys. Lett. B} {\bf 739}, 50 (2014).
\bibitem{ratbay} R. Myrzakulov, S. Odintsov and L. Sebastiani, arXiv:1412.1073 [gr-qc], (2014).
\bibitem{lorenzo1}K. Bamba, R. Myrzakulov,
S. D. Odintsov and L. Sebastiani, {\it Phys. Rev. D} {\bf 90}, 043505 (2014).
\bibitem{lorenzo2} L. Sebastiani, G. Cognola, R. Myrzakulov, S. D. Odintsov and S. Zerbini, {\it Phys. Rev. D} {\bf 89}, 023518 (2014).
%
%
\bibitem{capcurve}
S. Capozziello, {\it Int. J. Mod. Phys. D} {\bf 11}  (2002) 483.

\bibitem{Barth}N.H. Barth, S. Christensen, {\it Phys. Rev. D} {\bf 28}, 8 (1983).
%
\bibitem{nesseris}
  C.~Bogdanos, S.~Capozziello, M.~De Laurentis and S.~Nesseris,
  Astropart.\ Phys.\  {\bf 34}, 236  (2010).
%
\bibitem{paolella}
  M.~De Laurentis, M.~Paolella and S.~Capozziello,
  Phys.\ Rev.\ D {\bf 91},  083531 (2015).
  \bibitem{OdiGB} S. Nojiri and S.D. Odintsov, Phys.Lett. B {\bf 631}, 1, (2005). 
%
\bibitem{defelice}A. De Felice and T. Tanaka, {\it Prog. Theor. Phys.} {\bf 124}, 503 (2010). 
\bibitem{defelice1} A. De Felice and T. Suyama, {\it JCAP} {\bf 0906},  (2009) 034.
%


\bibitem{fGBnoether}S.~Capozziello, M.~De Laurentis and S.~D.~Odintsov, {\it Mod. Phys. Lett. A} {\bf 29}, 1450164 (2014).
%
\bibitem{antonio1} M. De Laurentis and A. J. Lopez-Revelles  {\it Int. J. Geom. Meth. Mod. Phys.} {\bf 11}, 1450082 (2014).
\bibitem{topoquint}M. De Laurentis,  {\it Mod. Phys. Lett. A } {\bf 30} , 1550069 (2015).
\bibitem{diego} E. Elizalde, R. Myrzakulov, V. V. Obukhov and D. S\'{a}ez-G\'{o}mez, {\it Class. Quant. Grav.} {\bf 27}, 095007 (2010).
\bibitem{myrz} R.~Myrzakulov, D.~S\'aez-G\'omez and A.~Tureanu, {\it Gen.\ Rel.\ Grav.} {\bf 43} ,1671 (2011).

%
\bibitem{DeLaurentis:2014oja} 
  M.~De Laurentis,
  Mod.\ Phys.\ Lett.\ A {\bf 30}, no. 12, 1550069 (2015).
  
 \bibitem{waves}
 B. Abbott et al. (Virgo, LIGO Scientific), Phys. Rev.
Lett. {\bf 119}, 161101 (2017).  
 
 \bibitem{ezquiaga}
  J.~M.~Ezquiaga and M.~Zumalacárregui,
  Phys.\ Rev.\ Lett.\  {\bf 119} (2017)   251304 (2017).
  
  
\bibitem{sasaki}
S. Nojiri, S. D. Odintsov, M. Sasaki,  Phys.Rev. D {\bf 71}, 123509  (2005).

\bibitem{Aghanim:2015xee} 
  N.~Aghanim {\it et al.} [Planck Collaboration],
  Astron.\ Astrophys.\  {\bf 594}, A11 (2016).


\bibitem{Betoule:2014frx} 
  M.~Betoule {\it et al.} [SDSS Collaboration],
  Astron.\ Astrophys.\  {\bf 568}, A22 (2014).
  
\bibitem{Riess} A. G. Riess {\it et al.}, 
ApJ {\bf 826}, 56 (2016).

\bibitem{Beutler:2013yhm} 
  F.~Beutler {\it et al.} [BOSS Collaboration],
  Mon.\ Not.\ Roy.\ Astron.\ Soc.\  {\bf 443}, no. 2, 1065 (2014).

\bibitem{camb}   
A.  Lewis,  A.  Challinor,  and  A.  Lasenby,  
Astrophys.  J. 538, 473 (2000).


\bibitem{cosmomc} A. Lewis and S. Bridle, 
Phys. Rev. D 66, 103511 (2002).

\bibitem{Ade:2015xua} 
  P.~A.~R.~Ade {\it et al.} [Planck Collaboration],
  Astron.\ Astrophys.\  {\bf 594}, A13 (2016).
    
    \bibitem{Trotta}
  R.~Trotta,
  Mon.\ Not.\ Roy.\ Astron.\ Soc.\  {\bf 378}, 72 (2007)
  
\bibitem{Trotta:2017wnx} 
  R.~Trotta,
  arXiv:1701.01467 [astro-ph.CO].
  
\bibitem{antonio}
  S.~Capozziello, S.~Nojiri, S.~D.~Odintsov and A.~Troisi,
  Phys.\ Lett.\ B {\bf 639}, 135  (2006).
  
\bibitem{sante} 
S. Carloni, P. Dunsby, S. Capozziello, A. Troisi, Class.
Quant. Grav. {\bf 22}, 4839 (2005).  
  
   
    
  \bibitem{Simony}
  S.~Santos da Costa, F. V. Roig , J. S. Alcaniz , S. Capozziello, M. De Laurentis and M. Benetti, submitted to Class. Quant. Grav.

\end{thebibliography}
\end{document}